\documentclass[11pt,a4paper]{article}

\usepackage[a4paper,hmargin=2.8cm,vmargin=3cm]{geometry}
 \usepackage[utf8x]{inputenc}
\usepackage{amsmath,amsfonts,amssymb,amsthm}
\usepackage{authblk} 
\usepackage[bookmarksnumbered=true]{hyperref}
\usepackage[capitalize]{cleveref}
\usepackage{graphicx}
\usepackage{tikz-cd}
\usetikzlibrary{positioning,calc,fadings,decorations.pathreplacing,shadings}

\newcommand\pgfmathsinandcos[3]{%
  \pgfmathsetmacro#1{sin(#3)}%
  \pgfmathsetmacro#2{cos(#3)}%
}
\newcommand\LongitudePlane[3][current plane]{%
  \pgfmathsinandcos\sinEl\cosEl{#2} 
  \pgfmathsinandcos\sint\cost{#3} 
  \tikzset{#1/.style={cm={\cost,\sint*\sinEl,0,\cosEl,(0,0)}}}
}

\newcommand\LatitudePlane[3][current plane]{%
  \pgfmathsinandcos\sinEl\cosEl{#2} 
  \pgfmathsinandcos\sint\cost{#3} 
  \pgfmathsetmacro\yshift{\RadiusSphere*\cosEl*\sint}
  \tikzset{#1/.style={cm={\cost,0,0,\cost*\sinEl,(0,\yshift)}}} %
}

\newcommand\DrawLongitudeArc[4][black]{
  \LongitudePlane{\angEl}{#2}
  \tikzset{current plane/.prefix style={scale=1}}
  \pgfmathsetmacro\angVis{atan(sin(#2)*cos(\angEl)/sin(\angEl))} %
  \pgfmathsetmacro\angA{mod(max(\angVis,#3),360)} %
  \pgfmathsetmacro\angB{mod(min(\angVis+180,#4),360} %
  \draw[current plane,#1]  (\angA:\RadiusSphere) arc (\angA:\angB:\RadiusSphere);
}%
\newcommand\DrawLatitudeCircle[2][1]{
  \LatitudePlane{\angEl}{#2}
  \tikzset{current plane/.prefix style={scale=1}}
  \pgfmathsetmacro\sinVis{sin(#2)/cos(#2)*sin(\angEl)/cos(\angEl)}
  \pgfmathsetmacro\angVis{asin(min(1,max(\sinVis,-1)))}
  \draw[current plane] (\angVis:\RadiusSphere) arc (\angVis:-\angVis-180:\RadiusSphere);
}

\newcommand\DrawLatitudeArc[4][black]{
  \LatitudePlane{\angEl}{#2}
  \tikzset{current plane/.prefix style={scale=1}}
  \pgfmathsetmacro\sinVis{sin(#2)/cos(#2)*sin(\angEl)/cos(\angEl)}
  \pgfmathsetmacro\angVis{asin(min(1,max(\sinVis,-1)))}
  \pgfmathsetmacro\angA{max(min(\angVis,#3),-\angVis-180)} %
  \pgfmathsetmacro\angB{min(\angVis,#4)} %

  \draw[current plane,#1] (\angA:\RadiusSphere) arc (\angA:\angB:\RadiusSphere);
}

\let\oldmarginpar\marginpar
\renewcommand\marginpar[1]{\-\oldmarginpar[\raggedleft\footnotesize #1]%
  {\raggedright\footnotesize #1}}

 \newtheorem{thm}{Theorem}[section]  

\renewcommand{\theequation}{\thesection.\arabic{equation}}

\newcommand{\dbtilde}[1]{\tilde{\raisebox{0pt}[0.85\height]{$\tilde{#1}$}}}

\newcommand{\di}{{\textrm{d}}}		

\newcommand{\Tbb}{\mathbb{T}}
\newcommand{\W}{W}
\newcommand{\Wt}{\tilde W}
\newcommand{\D}{D}

\newcommand{\Ecal}{\mathcal{E}}

\newcommand{\Ncal}{\mathcal{N}}		
\newcommand{\Hcal}{\mathcal{H}}		
\newcommand{\Ical}{\mathcal{I}}
\newcommand{\Ikp}{\Ical_{k}^{+}}

\newcommand{\Ik}{\Ical_{k}}

\newcommand{\Ocal}{\mathcal{O}}		
\newcommand{\hc}{\mbox{h.c.}}		

\newcommand{\Rbb}{\mathbb{R}}		
\newcommand{\Cbb}{\mathbb{C}}		
\newcommand{\Nbb}{\mathbb{N}}		
\newcommand{\Zbb}{\mathbb{Z}}

\newcommand{\id}{\mathbb{I}}
\newcommand{\norm}[1]{\lVert#1\rVert}	
\newcommand{\tr}{\operatorname{tr}}

\newcommand{\sgn}{\operatorname{sgn}}
\newcommand{\tagg}[1]{ \stepcounter{equation} \tag{\theequation} \label{eq:#1} } 

\newcommand{\north}{\Gamma^{\text{nor}}}

\newcommand{\diag}{\operatorname{diag}}
\newcommand{\diam}{\operatorname{diam}}
\newcommand{\supp}{\operatorname{supp}}

\title{Bosonic Collective Excitations in Fermi Gases}

\author[]{Niels Benedikter}

\affil[]{Universit\`a degli Studi di Milano, Dipartimento di Matematica, Via Cesare  Saldini 50, 20133 Milano, Italy\\\href{mailto:niels.benedikter@unimi.it}{niels.benedikter@unimi.it}}

\begin{document}





\maketitle   
\begin{abstract} 
Hartree--Fock theory has been justified as a mean--field approximation for fermionic systems. However, it suffers from some defects in predicting physical properties, making necessary a theory of quantum correlations. Recently, bosonization of many--body correlations has been rigorously justified as an upper bound on the correlation energy at high density with weak interactions. We review the bosonic approximation, deriving an effective Hamiltonian. We then show that for systems with Coulomb interaction this effective theory predicts collective excitations (plasmons) in accordance with the random phase approximation of Bohm and Pines, and with experimental observation.
 \end{abstract}
 



\section{Introduction}   
Hartree--Fock (HF) theory is a widely used approximation for fermionic many--body systems such as metals or atomic nuclei. It is particularly successful in mean--field parameter regimes, i.\,e., at high density and weak interaction. The HF approximation has been rigorously established for the time evolution of reduced density matrices \cite{BPS14b,BPS14a,BPS16,BSS18} and the ground state energy \cite{GS94}. However, HF theory suffers from defects such as predicting a vanishing density of states at the Fermi momentum, in contradiction to measurements of the specific heat. Recently, rigorous results going beyond HF theory have been obtained, estimating the correlation energy: as upper and lower bound to second order in perturbation theory \cite{HPR20} and as an upper bound including all orders of perturbation theory \cite{BNP+19} reproducing the predictions of \cite{Mac50,GB57}. The latter upper bound is based on bosonization of collective excitations, clarifying the nature of bosonic quasiparticles predicted by \cite{BP53,SBFB57}. Unlike, e.\,g., the Holstein--Primakoff bosonization \cite{HP40,CG12,CGS15,Ben17} or one--dimensional bosonization \cite{ML65}, collective bosonization is not an exact mapping but an approximation that requires estimates of its validity. In this paper we review three--dimensional bosonization and heuristically explore the predictions it makes about the excitation spectrum, in particular the emergence of plasma oscillations if a Coulomb interaction is present.
 
\medskip

We consider a system of a large number $N$ of fermions on the torus $\Tbb^3 = \Rbb^3/(2\pi\Zbb^3)$, interacting via a two--body potential $V$, described by the Hamiltonian
\begin{equation}\label{eq:firstquantop}H_N = -\hbar^2 \sum_{i=1}^N \Delta_{x_i} + \frac{1}{N}\sum_{1\leq i<j \leq N} V(x_i - x_j), \qquad \hbar := N^{-1/3}\;,\end{equation}
on the Hilbert space $L^2_a\left((\Tbb^3)^N\right)$ of functions that are antisymmetric under permutation of the $N$ arguments. The effective Planck constant $\hbar = N^{-1/3}$ and the coupling constant $N^{-1}$ model a high density regime with weak interactions (see \cite{BPS16} for an introduction). 
The ground state energy is defined as
\begin{equation}\label{eq:gse-HN}E_N := \inf_{\substack{\psi \in L^2_a((\Tbb^{3})^N)\\\norm{\psi} = 1}} \langle \psi, H_N \psi \rangle = \inf \operatorname{spec}\left( H_N \right)\;.\end{equation}
In Hartree-Fock theory, we restrict attention to Slater determinants
\begin{equation} \psi_\text{Slater} (x_1, \dots , x_N) = \frac{1}{\sqrt{N!}} \sum_{\sigma \in S_N} \sgn(\sigma) f_1 (x_{\sigma(1)}) f_2 (x_{\sigma(2)}) \dots f_N (x_{\sigma(N)}) \end{equation} with  $\{f_j\}_{j=1}^N$ a collection of orthonormal functions in $L^2 (\Tbb^3)$. The corresponding one--particle reduced density matrix is the projection operator $\omega = \sum_{j=1}^N \lvert f_j \rangle \langle f_j\rvert$. The infimum (over all rank-$N$ orthogonal projections $\omega$) of the HF functional
\begin{equation}
\begin{split}& \mathcal{E}_\text{HF}(\omega) := \langle \psi_\text{Slater}, H_N \psi_\text{Slater} \rangle = \tr \left(\!- \hbar^2 \Delta  \omega\right) \\
& \hspace{2.5em} +\! \frac{1}{N}\! \int\! \di x \di y V(x-y) \omega (x,x) \omega (y,y)-\! \frac{1}{N}\! \int\! \di x \di y V(x-y) \lvert \omega (x,y) \rvert^2\end{split}
\end{equation} 
provides a good approximation \cite{GS94} to $E_N$. The  minimizer of $\Ecal_\textnormal{HF}$ is hard to characterize, but according to \cite{GHL19}, the infimum only differs by an exponentially small amount from the energy we find using the $N$ plane waves
\begin{equation}\label{eq:plane-wave} f_{k_j}(x) = (2\pi)^{-3/2} e^{i k_j\cdot x}, \quad k_j \in \Zbb^3, \end{equation}
which minimize the kinetic energy $\tr (- \hbar^2 \Delta  \omega) = \hbar^2\sum_{j=1}^N \lvert k_j\rvert^2$.  In momentum space, the minimizing selection of momenta $k_j$ can be visualized as the Fermi ball 
\begin{equation}
B_\textnormal{F} := \{k \in \Zbb^3 : \lvert k \rvert \leq k_\textnormal{F}\}\;,
\end{equation} 
a ball around the origin of radius $k_\textnormal{F}$ chosen such that it contains $N$ points of $\Zbb^3$; i.\,e., $k_\textnormal{F} = (\frac{3}{4\pi})^{1/3}N^{1/3}$ up to lower order corrections. We write $\kappa := (\frac{3}{4\pi})^{1/3}$.

\medskip

 Obviously the infimum of $\Ecal_\textnormal{HF}$ is an upper bound to $E_N$. The difference between $E_N$ and $E_\textnormal{HF} := \inf_{\omega} \Ecal_\textnormal{HF}(\omega)$ is called correlation energy. In physics it is commonly calculated by partial resummation of diagrammatic perturbation theory \cite{Mac50,GB57}, going by the name of \emph{random phase approximation}. The perturbative prediction has recently been proven \cite{BNP+19} to be an upper bound for $E_N$ in systems with regular interaction potential; see the following theorem for the precise statement. 
 \begin{thm}[Random Phase Approximation as Upper Bound \cite{BNP+19}]\label{thm:main}
 Let $\hat{V}: \Zbb^3 \to \Rbb$ non--negative and compactly supported. Let the number of particles be $N := \lvert \{ k \in \Zbb^3 : \lvert k\rvert \leq k_\textnormal{F} \}\rvert$. Then for $k_\textnormal{F} \to \infty$ we have
 \begin{align*}
 & \frac{E_N - E_\textnormal{HF}}{\hbar} \leq  \tagg{upper} \\
 & \kappa \sum_{k \in \Zbb^3} \lvert k\rvert \Bigg[ \frac{1}{\pi}\! \int_0^\infty \!\!\!\!\!\log\Big( 1 + \hat{V}(k) \kappa 2\pi \Big(1 - \lambda \arctan \frac{1}{\lambda} \Big)\Big)\di\lambda - \hat{V}(k) \frac{\kappa \pi}{2} \Bigg] + \Ocal(N^{-1/27})\;.
 \end{align*}
 \end{thm}    
 A matching lower bound was proven recently \cite{BNP+20}. The proof of \cref{thm:main} is based on an effective bosonic theory which we review in \cref{sec:derivation}. In \cref{sec:energy} we discuss the predictions the effective theory makes about the excitation spectrum.
 
 \section{Derivation of the Bosonic Effective Hamiltonian}
 \label{sec:derivation} 
%
It is convenient to embed the system in fermionic Fock space and use creation and annihilation operators, $a^*_k$ and $a_k$, where the momenta $k$ refer to the plane wave basis, defined as in \cref{eq:plane-wave}. They satisfy the canonical anticommutator relations
\begin{equation}
\{a_k, a_l\} = 0 = \{a^*_k, a^*_l\}\;, \qquad \{a_k,a^*_l\} = \delta_{k,l}\;.
\end{equation}
The fermionic number operator is $\Ncal := \sum_{k \in \Zbb^3} a^*_k a_k$.
The Hamiltonian becomes
\begin{equation}
\Hcal_N = \sum_{k \in \Zbb^3} \hbar^2 \lvert k\rvert^2 a^*_k a_k + \frac{1}{2N}\sum_{k_1,k_2,k \in \Zbb^3} \hat{V}(k) a^*_{k_1} a^*_{k_2} a_{k_2+k} a_{k_1 -k}\;;
\end{equation}
more precisely, $\Hcal_N$ restricted to the eigenspace $\{\psi: \Ncal \psi = N\psi\}$ agrees with $H_N$.

\subsection{Particle--Hole Transformation}
We start with a particle--hole transformation which extracts the HF energy and leaves us with a remainder Hamiltonian describing quantum correlations. This transformation is a unitary $R$ on fermionic Fock space defined by its action
\begin{equation}\label{eq:phtrafo}
R^* a^*_k R = \left\{ \begin{array}{cl}
                     a^*_k & \textnormal{for } k \in B_\textnormal{F}^c := \Zbb^3 \setminus B_\textnormal{F} \\ a_{-k} & \textnormal{for } k\in B_\textnormal{F} \end{array} \right.
\end{equation}
and by transforming the vacuum $\Omega := (1,0,0,0,\ldots)$ into the Slater determinant corresponding to the Fermi ball, $R\Omega = \bigwedge_{k \in B_\textnormal{F}} f_k$.
Applying \cref{eq:phtrafo} and then rearranging in normal order at the cost of anticommutators appearing, we find
\begin{equation}\label{eq:Htrafo}
R^* \Hcal_N R = \Ecal_\textnormal{HF}(\omega) + \sum_{p \in B_\textnormal{F}^c} \hbar^2\lvert p\rvert^2 a^*_p a_p - \sum_{h \in B_\textnormal{F}} \hbar^2 \lvert h\rvert^2 a^*_h a_h + Q_N + \mathcal{O}\left( \frac{\Ncal^2+1}{\sqrt{N}} \right)\;,
\end{equation}
where now $\omega= \sum_{k\in B_\textnormal{F}} \lvert f_k\rangle\langle f_k \rvert$.
The operator $Q_N$ is quartic in fermionic operators; it describes repulsion of particles with particles (momenta ``$p$'') and holes with holes (momenta ``$h$'') as well as attraction between particles and holes.
The last term is negligible in the approximate ground state constructed by \cite{BNP+19}
and in states with small number of excitations. 
The negative sign of the term $- \sum_{h \in B_\textnormal{F}} \hbar^2 \lvert h\rvert^2 a^*_h a_h$ can be understood as saying that creation of a hole in the Fermi ball lowers the energy.

\subsection{Introduction of Almost--Bosonic Quasiparticles}
We introduce pair excitation operators that lift a fermion from inside the Fermi ball to outside the Fermi ball by a relative momentum $k \in \Zbb^3$,  delocalized over all the Fermi surface, by
\begin{equation}
\tilde{b}^*_k := \sum_{\substack{p \in B_\textnormal{F}^c\\h \in B_\textnormal{F}}} \delta_{p-h,k} a^*_p a^*_h\;.
\end{equation}
In terms of these operators the interaction $Q_N$ appearing in \cref{eq:Htrafo} can be written
\begin{equation}
Q_N = \frac{1}{2N}\sum_{k \in \Zbb^3} \hat{V}(k) \left( 2 \tilde{b}^*_k \tilde{b}_k + \tilde{b}^*_k \tilde{b}^*_{-k} + \tilde{b}_{-k} \tilde{b}_k \right)\;.
\end{equation}
Just as for bosons $[\tilde{b}^*_k,\tilde{b}^*_l]=0$. Unfortunately this is not sufficient for the pair operators to satisfy canonical commutator relations: consider a quasiparticle created by $b^*_{p,h} := a^*_p a^*_h$ as proposed by \cite{SBFB57}, then obviously $(b^*_{p,h})^2 =0$: we cannot create more than one such quasiparticle, so it cannot be bosonic. However, $\tilde{b}^*_{k}$ creates a superposition delocalized over many fermionic modes; thus $\big(\tilde{b}^*_k\big)^r$ only vanishes once $r \in \Nbb$ becomes larger than the number of available fermionic modes. So we expect the operators $\tilde{b}^*_k$ to be approximately bosonic on states where the number of occupied fermionic modes is much smaller than the number of modes constituting the superposition. In fact 
\begin{equation}\label{eq:noccr}
[\tilde{b}_k,\tilde{b}^*_l] = \textnormal{const.}\times \left(\delta_{k,l} + \mathcal{E}(k,l) \right)
\end{equation}
where the operator $\mathcal{E}(k,l)$ is to be thought of as a small error. Controlling this type of error term is a central task achieved in \cite{BNP+19}. 
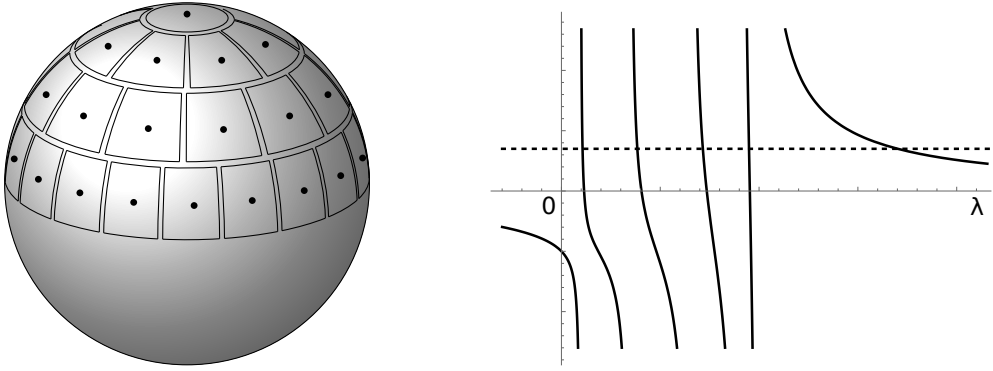
\begin{figure}\centering
 \begin{minipage}[t]{0.4\textwidth}\centering
\hspace{-1em}\begin{tikzpicture}[scale=0.6] 
\def\RadiusSphere{4} 
\def\angEl{20} 
\def\angAz{-20} 

\filldraw[ball color = white] (0,0) circle (\RadiusSphere);

\DrawLatitudeCircle[\RadiusSphere]{75+2}
\foreach \t in {0,-50,...,-250} {
  \DrawLatitudeArc{75}{(\t+50-4)*sin(62)}{\t*sin(62)}
 \DrawLongitudeArc{\t*sin(62)}{50+2}{75}
 \DrawLongitudeArc{(\t-4)*sin(62)}{50+2}{75}
  \DrawLatitudeArc{50+2}{(\t+50-4)*sin(62)}{\t*sin(62)}
 }
 \foreach \t in {0,-50,...,-300} {
   \DrawLatitudeArc{50}{(\t+50-4)*sin(37)}{\t*sin(37)}
 \DrawLongitudeArc{\t*sin(37)}{25+2}{50}
  \DrawLongitudeArc{(\t-4)*sin(37)}{25+2}{50}
   \DrawLatitudeArc{25+2}{(\t+50-4)*sin(37)}{\t*sin(37)}
 }
 \DrawLatitudeArc{50}{(-300-4)*sin(37)}{-330*sin(37)}
 \foreach \t in {0,-50,...,-450} {
    \DrawLatitudeArc{25}{(\t+50-4)*sin(23)}{\t*sin(23)}
 \DrawLongitudeArc{\t*sin(23)}{00+2}{25}
 \DrawLongitudeArc{(\t-4)*sin(23)}{00+2}{25}
 \DrawLatitudeArc{00+2}{(\t+50-4)*sin(23)}{\t*sin(23)}
 }
     \DrawLatitudeArc{25}{(-450-4)*sin(23)}{-500*sin(23)}

\fill[black] (0,3.75) circle (.075cm);

\fill[black] (1.72,3.08) circle (.075cm);
\fill[black] (.76,2.73) circle (.075cm);
\fill[black] (-.66,2.73) circle (.075cm);
\fill[black] (-1.73,3.04) circle (.075cm);

\fill[black] (2.25,1.5) circle (.075cm);
\fill[black] (.8,1.2) circle (.075cm);
\fill[black] (-.85,1.22) circle (.075cm);
\fill[black] (-2.27,1.5) circle (.075cm);
\fill[black] (-3.09,1.97) circle (.075cm);
\fill[black] (3.09,1.97) circle (.075cm);

\fill[black] (2.57,-.15) circle (.075cm);
\fill[black] (1.43,-.37) circle (.075cm);
\fill[black] (.155,-.48) circle (.075cm);
\fill[black] (-1.17,-.41) circle (.075cm);
\fill[black] (-2.35,-.2) circle (.075cm);
\fill[black] (-3.26,0.1) circle (.075cm);
\fill[black] (-3.79,.55) circle (.075cm);
\fill[black] (3.37,.18) circle (.075cm);
\fill[black] (3.85,.57) circle (.075cm);

\end{tikzpicture}
\end{minipage}
 \begin{minipage}[t]{0.51\textwidth}\centering
 \includegraphics[scale=0.4]{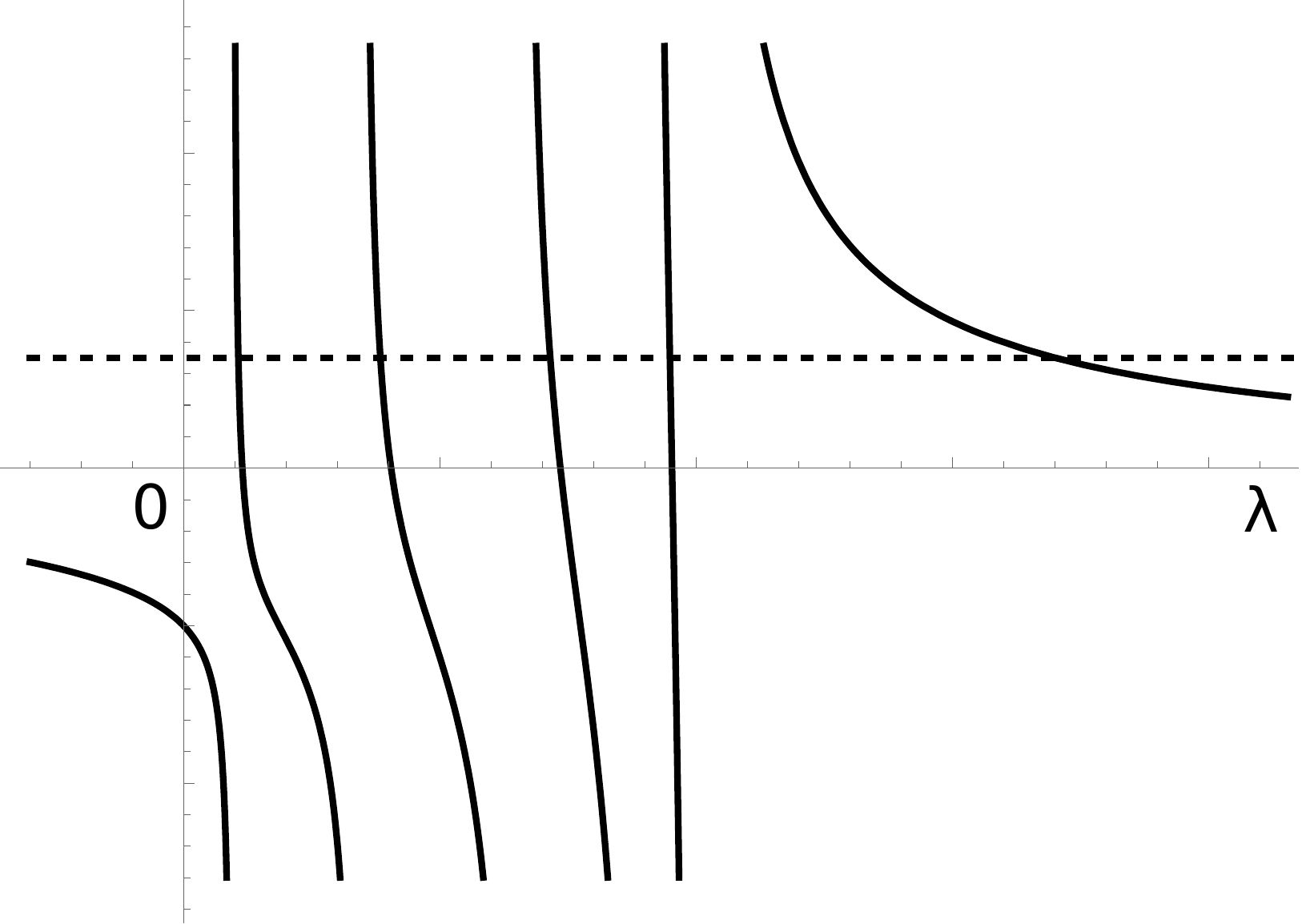} 
 \end{minipage}
 \caption{\textbf{Left:} Fermi surface decomposed into patches separated by corridors of width $2R$, $R := \diam\supp\hat{V}$. The area covered by a patch is approximately $4\pi N^{2/3}/M$. Extending the $\alpha$--th patch radially in- and outward by $R$ defines $B_\alpha \subset \Zbb^3$. The vectors ${\omega}_\alpha$ are the centers of the patches, marked by dots. Patches are reflected across the center to the southern half sphere. \textbf{Right:}
    Left hand side of \cref{eq:tosolve} as function of $\lambda$ as solid line (poles at unperturbed eigenvalues $\lvert \hat{k}\cdot \hat{\omega}_\alpha \rvert^2$), right hand side dashed. Perturbed eigenvalues are the $\lambda$ where curves intersect. The perturbed eigenvalues interlace the unperturbed eigenvalues. As $\lvert k\rvert \to 0$, the dashed line moves toward the horizontal axis and the largest eigenvalue  toward $+ \infty$, giving rise to the plasmon.}   
\label{fig:blub}
\end{figure}


However, another problem appears: the kinetic energy cannot be expressed by such operators. To solve this problem we cut the Fermi surface into patches $B_\alpha$ as sketched in \cref{fig:blub} on the left and localize the pair operators accordingly. It turns out that the number of fermionic modes per patch is still large enough to justify the bosonic approximation if the number of patches is $M \ll N^{2/3}$. So let
\begin{equation} \label{eq:locbos}
b^*_{\alpha,k} := \frac{1}{n_{\alpha,k}} \sum_{\substack{p \in B_\textnormal{F}^c \cap B_\alpha\\h \in B_\textnormal{F} \cap B_\alpha}} \delta_{p-h,k} a^*_p a^*_h\;;
\end{equation}
the definition of the normalization constant $n_{\alpha,k}$ is discussed below. We write $\omega_\alpha$ for the vector pointing to the center of the patch $B_\alpha$. The operators $b^*_{\alpha,k}$ create approximate eigenmodes of the kinetic energy; more precisely, by using $\lvert p\rvert^2 - \lvert h\rvert^2 = \left(p-h\right)\cdot\left(p+h\right) \simeq k\cdot (2\omega_\alpha)$ we obtain the commutator
\begin{align}
\hbar^2\Bigg[\sum_{p \in B_\textnormal{F}^c} \lvert p\rvert^2 a^*_p a_p - \sum_{h \in B_\textnormal{F}} \lvert h\rvert^2 a^*_h a_h, b^*_{\alpha,k}\Bigg] & = \frac{\hbar^2}{n_{\alpha,k}} \sum_{\substack{p \in B_\textnormal{F}^c \cap B_\alpha\\h \in B_\textnormal{F} \cap B_\alpha}} \delta_{p-h,k}  \left( \lvert p\rvert^2 - \lvert h\rvert^2 \right) a^*_p a^*_h \nonumber \\
& = \hbar 2\kappa \,k \cdot\hat{\omega}_\alpha + \Ocal\left(\frac{\Ncal}{M}\right)\;.
\end{align}
If $M \gg N^{1/3}$, the error term is smaller than $\hbar$ and thus negligible. Notice that there is a subtlety: if $k\cdot \hat{\omega}_\alpha < 0$, the relative momentum $k$ points from outside the Fermi ball to inside, incompatible with the summation in \cref{eq:locbos} being such that $k$ points from a hole momentum $h\in B_\textnormal{F}$ to a particle momentum $p \in B_\textnormal{F}^c$. So in this case the sum in \cref{eq:locbos} is empty. Similarly, there may still be very few summands if $k\cdot\hat{\omega}_\alpha \simeq 0$. We thus impose a cutoff $\hat{k}\cdot\hat{\omega}_\alpha \geq N^{-\delta}$ (with a small $\delta >0$ to be optimized), and call the set of patch indices $\alpha$ satisfying this condition $\Ikp$. Since there are going to be interaction terms coupling $k$ to $-k$, it is convenient to also introduce the set $\north \subset \Zbb^3$, denoting an arbitrarily chosen half space. To conclude, as in the classical one-dimensional bosonization of the Luttinger model \cite{ML65}, we now propose to approximate the kinetic energy by
\begin{equation}\begin{split}
& \hbar^2 \sum_{p \in B_\textnormal{F}^c} \lvert p\rvert^2 a^*_p a_p - \hbar^2\sum_{h \in B_\textnormal{F}} \lvert h\rvert^2 a^*_h a_h \\
& \simeq \hbar 2\kappa\sum_{k \in \north} \lvert k\rvert\Bigg( \sum_{\alpha \in\Ikp} \lvert \hat{k}\cdot \hat{\omega}_\alpha\rvert b^*_{\alpha,k} b_{\alpha,k} + \sum_{\beta \in\Ical_{-k}^+} \lvert \hat{k}\cdot \hat{\omega}_\beta\rvert b^*_{\beta,-k} b_{\beta,-k} \Bigg)\;.
\end{split}
\end{equation}
Decomposing $\tilde{b}^*_k$ into the $b^*_{\alpha,k}$ we arrive at a quadratic, approximately bosonic, Hamiltonian compromising both kinetic and interaction energy. It only remains to determine the normalization constant $n_{\alpha,k}$; we fix it by imposing that the leading order of $[b_{\alpha,k},b^*_{\beta,l}]$ is given by $\delta_{k,l}\delta_{\alpha,\beta}$ as in the canonical commutator relations; this is achieved by $n_{\alpha,k}^2 := \sum_{\substack{p \in B_\textnormal{F}^c \cap B_\alpha\\h \in B_\textnormal{F} \cap B_\alpha}} \delta_{p-h,k}$, the number of pairs of relative momentum $k$ in patch $B_\alpha$. If $\hat{k}\cdot\hat{\omega}_{\alpha} \simeq 1$ this can be identified with the volume of a flat box over the Fermi surface having base area $4\pi N^{2/3} M^{-1}$ and height $\lvert k\rvert$. As mentioned before, the number of pairs $(p,h)$, $p \in B_\textnormal{F}^c \cap B_\alpha$ and $h \in B_\textnormal{F} \cap B_\alpha$ with $p-h=k$, is small when $k \cdot \omega_\alpha \simeq 0$ (only very few ``tangential'' particle--hole excitation are possible). The correct interpolation between these extreme cases is found to be
$n_{\alpha,k}^2 \simeq {4\pi N^{2/3}} M^{-1} \lvert k\cdot\hat{\omega}_\alpha \rvert$.
We thus find the effective Hamiltonian
\begin{equation}\label{eq:directsum}H_\text{eff} = \hbar 2\kappa \sum_{k \in \north} \lvert k\rvert h_\text{eff}(k)\end{equation} 
where, with ${u_\alpha}(k) := \sqrt{\lvert \hat{k}\cdot\hat{\omega}_{\alpha}\rvert}$, we have
\begin{align}h_\text{eff}(k) & = \sum_{\alpha \in\Ikp} u_\alpha(k)^2 b^*_{\alpha,k} b_{\alpha,k} + \sum_{\alpha \in\Ical_{-k}^{+}} u_\alpha(k)^2 b^*_{\alpha,-k} b_{\alpha,-k} \nonumber\\
& \quad + \frac{\kappa \hat{V}(k) 2\pi}{M} \Big( \sum_{\alpha,\beta \in \Ikp}\! u_\alpha(k) u_\beta(k) b^*_{\alpha,k} {b}_{\beta,k} +\! \sum_{\alpha,\beta \in\Ical_{-k}^{+}} \! u_\alpha(k) u_\beta(k) b^*_{\alpha,-k} b_{\beta,-k} \nonumber \\
&\hspace{2.8cm} + \bigg[ \sum_{\alpha \in \Ikp}\sum_{\beta \in \Ical_{-k}^{+}} u_\alpha(k) u_\beta(k) b^*_{\alpha,k} b^*_{\beta,-k} + \hc\bigg]\Big).\end{align}
 We think of $H_\text{eff}$ as a refinement of the Hamiltonian proposed by Sawada et al.~\cite{SBFB57}.
%
%

\section{Excitation Spectrum of the Bosonic Effective Hamiltonian}\label{sec:energy}
\emph{For the following, we use the approximation (compare to \cref{eq:noccr}) that the $b$-- and $b^*$--operators exactly satisfy canonical commutator relations, i.\,e.,}
\begin{equation}\label{eq:ccr}
[b_{\alpha,k}, b_{\beta,l}] = 0 = [b^*_{\alpha,k}, b^*_{\beta,l}]\;, \qquad [b_{\alpha,k}, b^*_{\beta,l}] = \delta_{\alpha,\beta} \delta_{k,l}\;.   
\end{equation}
The approximation only lies in the last relation; we gave a rigorous estimate for the error in \cite[Lemma 4.1]{BNP+19}, showing that it is small if there are only few fermionic excitations over the Fermi ball. In \cite[Proposition 4.6]{BNP+19} we proved that the bosonic approximation to the ground state of \cref{eq:directsum} contains only few fermionic excitations, and so do also states obtained by adding a fixed number of bosonic excitations to the approximate ground state, i.\,e., by applying $b^*_{\alpha,k}$--operators.

\medskip
 
With this approximation, operators belonging to different $k$ commute, so we can diagonalize every $h_\text{eff}(k)$ independently and in the end sum over $k \in \north$. In the following we mostly drop the $k$--dependence in the notation . 
Keeping only the non--vanishing operators, we set
\begin{equation}\label{eq:coperators}c^*_\alpha := \left\{ \begin{array}{lr} b^*_{\alpha,k} & \text{ for } \alpha \in \Ikp\,, \\ b^*_{\alpha,-k} & \text{ for } \alpha \in 
\Ical_{-k}^{+}\;. \end{array}\right.\end{equation}
These operators again satisfy canonical commutator relations, $[c_\alpha,c_\beta] = 0 = [c^*_\alpha,c^*_\beta]$ and $[c_\alpha,c^*_\beta] =\delta_{\alpha,\beta}$. We also introduce the index set $\Ik := \Ikp \cup \Ical_{-k}^{+}$ and let $I_k := \lvert \Ical_k^{+}\rvert = \lvert \Ical_{-k}^{+}\rvert$. As in \cite{GS13} we write
\begin{equation}\label{eq:hameff}
h_\textnormal{eff} = \mathbb{H} - \frac{1}{2} \tr (\D+\W)\,,
\end{equation}
\begin{equation}\mathbb{H} := \frac{1}{2} \begin{pmatrix} (c^*)^T & c^T \end{pmatrix} \begin{pmatrix} \D+\W & \Wt \\ \Wt & \D+\W\end{pmatrix} \begin{pmatrix}c\\c^* \end{pmatrix},\quad  c = \begin{pmatrix} \vdots \\ c_\alpha \\ \vdots\end{pmatrix},\quad c^* = \begin{pmatrix} \vdots \\ c^*_\alpha \\ \vdots \end {pmatrix},\end{equation}
where $c^T = \begin{pmatrix} \cdots & c_\alpha & \cdots \end{pmatrix}$.
Setting $g :=  \kappa \hat{V}2 \pi/M$, the matrices $\D$, $\W$, and $\Wt$ are 
\begin{equation}\label{eq:blocks}\begin{split}
D & := \diag(u_\alpha^2: \alpha \in \Ik)\;, \\
\W_{\alpha,\beta} & := \left\{ \begin{array}{cl} g u_\alpha u_\beta \quad & \text{for } \alpha,\beta \in \Ikp \textnormal{ or } \alpha,\beta \in \Ical_{-k}^{+} \\ 0 & \text{for } \alpha \in \Ikp, \beta \in \Ical_{-k}^{+} \text{ or } \alpha \in \Ical_{-k}^{+}, \beta \in \Ikp\;,\end{array} \right. \\
\Wt_{\alpha,\beta} & := \left\{ \begin{array}{cl} 0 & \text{for } \alpha,\beta \in \Ikp \text{ or } \alpha,\beta \in \Ical_{-k}^{+} \\
g u_\alpha u_\beta \quad & \text{for } \alpha \in \Ikp, \beta \in \Ical_{-k}^{+} \textnormal{ or } \alpha \in \Ical_{-k}^{+},\beta \in \Ikp\;.\end{array} \right.
\end{split}\end{equation}
By reordering the indices we can write
\begin{equation}D = \begin{pmatrix} d & 0 \\ 0 & d \end{pmatrix}, \quad W = \begin{pmatrix} b & 0 \\ 0 & b \end{pmatrix}, \quad \tilde{W} = \begin{pmatrix} 0 & b \\ b& 0\end{pmatrix}\end{equation}
where $d = \diag(u_\alpha^2: \alpha=1,\ldots I_k)$, $b = g\lvert u\rangle \langle u\rvert$, and $u = ( u_1 ,\ldots, u_{I_k} )^T$.
The Segal field operators $\phi = \begin{pmatrix} \cdots & \phi_\alpha & \cdots \end{pmatrix}^T$ and $\pi = \begin{pmatrix} \cdots & \pi_\alpha & \cdots \end{pmatrix}^T$ are defined by
\begin{equation}\label{eq:defsegalfields}\begin{pmatrix}
   c \\ c^*
  \end{pmatrix} = \Theta \begin{pmatrix} \phi \\ \pi\end{pmatrix}, \quad \textnormal{where } \Theta := \frac{1}{\sqrt{2}} \begin{pmatrix} 1 & i\\ 1& -i \end{pmatrix}.
\end{equation}
Note that $\phi = \frac{1}{\sqrt{2}}(c+c^*) = \phi^*$ and $\pi = \frac{i}{\sqrt{2}}(c^*-c) = \pi^*$. Then
\begin{equation}\label{eq:quadratichamiltonian}\begin{split}\mathbb{H} & = \begin{pmatrix}\phi^T & \pi^T \end{pmatrix} \mathfrak{M} \begin{pmatrix} \phi \\ \pi\end{pmatrix}, \textnormal{ with } \mathfrak{M} = \frac{1}{2} \begin{pmatrix}                                                                                                                                                                                                    \D+\W+\Wt & 0\\ 0 & \D+\W-\Wt                                                                                                                                                                            \end{pmatrix}\;.\end{split}
\end{equation}
The commutator relations of the Segal field operators are invariant under symplectic transformations, which corresponds to Bogoliubov transformations of the bosonic creation and annihilation operators. We introduce
\begin{equation}E := \left( (\D+\W-\Wt)^{1/2} (\D+\W+\Wt) (\D+\W-\Wt)^{1/2} \right)^{1/2} \in \Cbb^{2I_k\times 2I_k}\end{equation}
and, with $\id$ denoting the $I_k\times I_k$--identity matrix, the symplectic matrix $S$, 
\begin{equation}\label{eq:defS}S  := \begin{pmatrix} (\D+\W-\Wt)^{1/2} E^{-1/2} U & 0 \\ 0 & (\D+\W-\Wt)^{-1/2} E^{1/2} U \end{pmatrix}, \quad 
U := \frac{1}{\sqrt{2}}\begin{pmatrix} \id & \id \\ \id & -\id \end{pmatrix}.\end{equation}
The matrix $U$ block--diagonalizes $D+W-\tilde{W}$ and $D+W+\tilde{W}$.
We then find
\begin{equation}\label{eq:completediag}\mathbb{H}  = \begin{pmatrix}\tilde\phi^T & \tilde\pi^T \end{pmatrix} \frac{1}{2}\begin{pmatrix} \tilde{E} & 0\\ 0 & \tilde{E}\end{pmatrix} \begin{pmatrix} \tilde\phi \\ \tilde\pi\end{pmatrix}\;,\quad \textnormal{having set } \begin{pmatrix} \tilde{\phi} \\ \tilde{\pi} \end{pmatrix} := S^{-1} \begin{pmatrix}\phi \\ \pi \end{pmatrix}\end{equation}
and having introduced the matrix
\begin{equation}\tilde{E} = \begin{pmatrix} \left[ d^{1/2} (d+2b) d^{1/2}\right]^{1/2} & 0 \\ 0 & \left[ (d+2b)^{1/2}d(d+2b)^{1/2} \right]^{1/2} \end{pmatrix} =: \begin{pmatrix} A^{1/2} & 0 \\ 0& B^{1/2} \end{pmatrix}\;.\label{eq:tildeE}\end{equation}
Since $\tilde{E}$ is symmetric we can find an orthogonal matrix $O$ such that $O^T \tilde{E} O = \diag(e_\gamma: \gamma \in \Ik)$.
After the further symplectic transformation $\tilde{S} := \begin{pmatrix} O & 0 \\ 0 & O \end{pmatrix}$ then
\begin{equation}\label{eq:diagonal}
\begin{split}\mathbb{H} & = \begin{pmatrix}\dbtilde\phi^T & \dbtilde\pi^T \end{pmatrix} \frac{1}{2}\begin{pmatrix} \diag(e_\gamma) & 0\\ 0 & \diag(e_\gamma)\end{pmatrix} \begin{pmatrix} \dbtilde\phi \\ \dbtilde\pi\end{pmatrix} \\
& = \sum_{\gamma \in \Ik} \frac{e_\gamma}{2} \left( \dbtilde{\phi}_\gamma^2 + \dbtilde{\pi}_\gamma^2 \right) = \sum_{\gamma \in \Ik} e_\gamma \left( \hat{n}_\gamma + \frac{1}{2} \right)\;,\end{split}\end{equation}
where we recognized harmonic oscillators and introduced the corresponding number operators $\hat{n}_\gamma$. In particular, \cref{eq:diagonal} implies $h_\textnormal{eff} \geq \frac{1}{2} \tr \left(E- (D+W)\right)$, which becomes \cref{eq:upper} as $M\to\infty$ \cite{BNP+19}. The total excitation spectrum is
\begin{equation}\sigma(H_\textnormal{eff}) = \Big\{ \hbar\kappa \sum_{k \in \north} \lvert k\rvert\Big[ \sum_{\gamma \in \Ik} 2 e_\gamma(k) n_\gamma(k) +  \tr\big(E(k)-D(k)-W(k)\big)\Big]: n_\gamma(k) \in \mathbb{N} \Big\}\;.\end{equation}
In the following we approximately compute the excitation energies $e_\gamma(k)$.

\subsection{The Plasmon Dispersion Relation}
Recall the definition of the matrices $A$ and $B$ from \cref{eq:tildeE}. We observe that $A = d^{1/2}(d+2b)d^{1/2} = d^2 + 2g \lvert \tilde u\rangle \langle \tilde u\rvert$ where $\tilde u := d^{1/2} u$ is the vector with components $u_\alpha^2$; i.\,e., $A$ is diagonal plus a rank--one perturbation. Define $X := (d+2b)^{-1/2} d^{-1/2}$; then $A = X^{-1}BX$ and thus $A$ and $B$ have the same spectrum. Thus, we only need to find the eigenvalues of $A$ (or more precisely, of $A^{1/2}$).

The spectrum of $A$ can be obtained by the following standard method: by the matrix determinant lemma the characteristic polynomial of $A$ can be written as
\begin{equation}\begin{split}& \det(d^2 + 2g \lvert \tilde{u}\rangle\langle \tilde{u}\rvert - \lambda) = \det(d^2-\lambda)\det\left(1 + 2g(d^2-\lambda)^{-1}\lvert \tilde{u}\rangle\langle \tilde{u}\rvert\right)\\
& = \prod_{\beta=1}^{I_k} (u_\beta^4-\lambda) \left( 1 + 2g \sum_{\alpha=1}^{I_k} \frac{u_\alpha^4}{u_\alpha^4 - \lambda} \right) =: \prod_{\beta=1}^{I_k} (u_\beta^4-\lambda) w(\lambda)\;.\end{split}\end{equation}
For $\lambda \to u_\alpha^4$ this expression is non--zero, thanks to the previously introduced cutoff $u_\alpha^2 \geq N^{-\delta}$. So the characteristic polynomial vanishes if and only if $w(\lambda)$ has a zero. 

\begin{figure}\centering
 \begin{minipage}[b]{0.48\textwidth}\centering
       \includegraphics[scale=0.55]{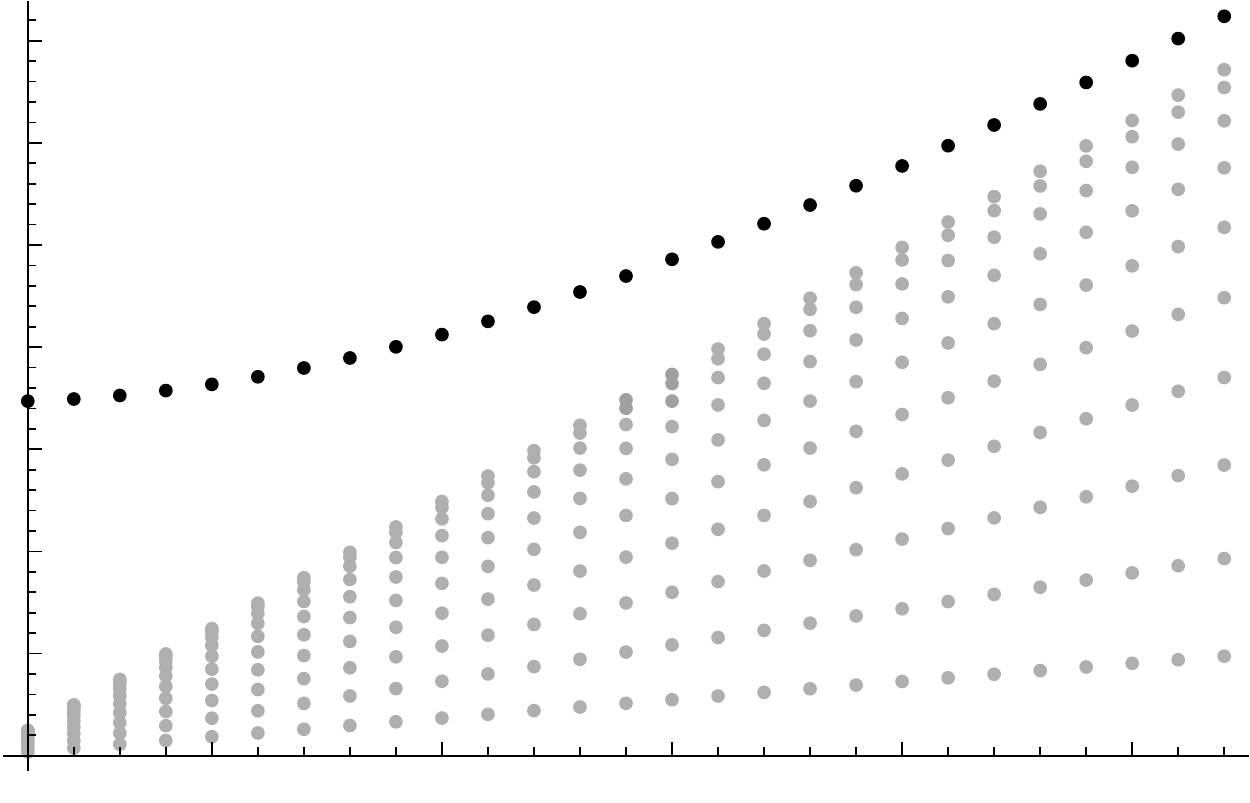} 
                \\
         \vspace{-1em}
         {\footnotesize \hspace{3.5em}Wavevector $k$ of particle--hole pair}
%
 \end{minipage}\hspace{1.0em}
 \begin{minipage}[b]{0.48\textwidth} 
 \centering
%
         \includegraphics[scale=0.62]{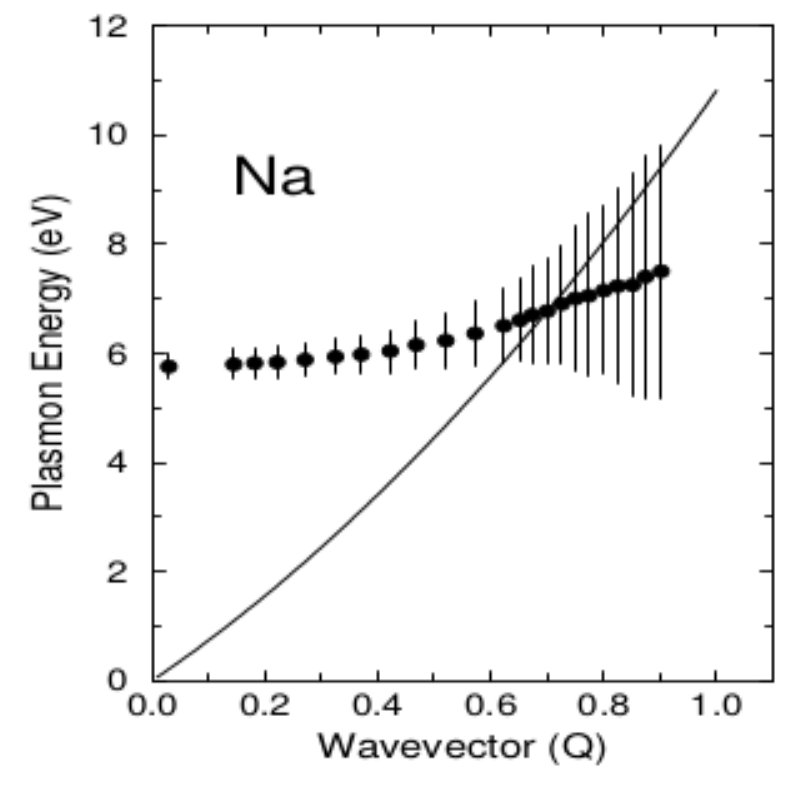}
 \end{minipage}
\caption[Comparison of Excitation Spectra]{
\textbf{Left:} Numerically determined excitation spectrum of $\hbar 2 \kappa \lvert k\rvert h_\textnormal{eff}(k)$ (ground state energy has been subtracted). Plasmon in black, grey points approximate continuum as $M \to \infty$.
\textbf{Right:} Excitation spectrum of sodium, plot from \cite{All96} based on data from \cite{vSF89}. The area below the solid line represents the continuum of excitations; black marks indicate the plasmon. 
}
\label{fig:monspectrum} 
\end{figure}

Considering the Coulomb potential $\hat{V}(k) = \lvert k\rvert^{-2}$, the condition $w(\lambda) =0$ means
\begin{equation}
\label{eq:tosolve}\frac{1}{M}\sum_{\alpha=1}^{2I_k} \frac{\lvert \hat{k}\cdot \hat{\omega}_\alpha \rvert^2}{\lambda - \lvert \hat{k}\cdot \hat{\omega}_\alpha \rvert^2} = \frac{\lvert k\rvert^2}{4\pi\kappa}\;.\end{equation}
 This has solutions that interlace the unperturbed eigenvalues $\lvert \hat{k}\cdot\hat{\omega}_\alpha \rvert^2$ and another solution at $\lambda > \max_\alpha \lvert \hat{k}\cdot\hat{\omega}_\alpha \rvert^2$ (see the right of \cref{fig:blub}); the latter solution will be identified as the plasmon. To calculate its energy, note that here the summand is free of singularities, and so, approximating the Riemann sum $\frac{4\pi}{M}\sum_{\alpha}$ by the surface integral over the unit 2--sphere, the left hand side becomes approximately
\begin{equation}\frac{1}{4\pi} \int_0^\pi \di\theta \sin \theta \int_0^{2\pi} \di\varphi \frac{\cos^2 \theta}{\lambda - \cos^2 \theta} = -1 + \sqrt{\lambda} \operatorname{arcoth}(\sqrt{\lambda})\;.\end{equation}
(We chose spherical coordinates such that $\hat{k}\cdot\hat{\omega}_\alpha = \cos\theta$.)
Solving \cref{eq:tosolve} for $\sqrt{\lambda}$ yields (approximately) the eigenvalues of $A^{1/2}$.  
As we are interested in large $\sqrt{\lambda}$, we use the series expansion $\operatorname{arcoth}(\sqrt{\lambda}) = \sqrt{\lambda}^{-1} + \frac{1}{3}\sqrt{\lambda}^{-3} + \frac{1}{5}\sqrt{\lambda}^{-5} + \Ocal(\sqrt{\lambda}^{-7})$ and then solve the power series for $\sqrt{\lambda}$, yielding
\begin{equation}\sqrt{\lambda} = \frac{2\sqrt{\pi\kappa}}{\lvert k\rvert \sqrt{3}} + \frac{3^{3/2}}{20} \frac{\lvert k\rvert}{\sqrt{\pi \kappa}} + \Ocal(\lvert k\rvert^3) \quad \textnormal{for small } k\;.\end{equation}
(For a precise notion of ``small $k$'' one may consider a large--volume limit or compare to a coupling strength parameter.) 
Multiplying by the prefactor $\hbar\kappa 2\lvert k\rvert$ from \eqref{eq:directsum}, and recalling $\kappa = \left(\frac{3}{4\pi}\right)^{1/3}$, the \emph{excitation energy (dispersion relation)} becomes
\begin{equation}\label{eq:plasmondisprel}
\lambda_\textnormal{pl}(k) = \hbar 2 + \hbar \frac{3}{10} \sqrt{\frac{3\kappa}{\pi}} \lvert k\rvert^2 + \Ocal(\lvert k\rvert^4)\;.
\end{equation}
The first summand, $\lambda_\textnormal{pl}(0) = \hbar 2$, is the frequency of classical plasma oscillations. But we also reproduce the quantum correction: in the physics literature \cite{vSF89} the plasmon is described as an excitation with dispersion relation 
\begin{equation}\label{eq:experi}
\lambda_\textnormal{pl}(k) = \lambda_\textnormal{pl}(0) + \frac{\hbar^2}{m} \alpha_\textnormal{RPA} \lvert k\rvert^2 + \Ocal(\lvert k\rvert^4)\;, \qquad \textnormal{with } \alpha_\textnormal{RPA} = \frac{3}{5} \frac{E_F}{\lambda_\textnormal{pl}(0)}\;.
\end{equation}
Since the Fermi energy is given by $E_F = \hbar^2 k_F^2$, the particle mass $m=1/2$, and $k_F = \kappa N^{1/3}$, we see that \cref{eq:plasmondisprel} and \cref{eq:experi} agree.
In \cref{fig:monspectrum} on the left we plotted all solutions of \cref{eq:tosolve} as a function of $\lvert k\rvert$; on the right we show experimental data for the excitation spectrum of sodium as plotted in \cite{All96}.

\section*{Acknowledgements}
NB has been supported by ERC grant 694227 and by Swedish Research Council grant 2016-06596 and Verg Foundation while in residence at Institut Mittag--Leffler.

\bibliographystyle{alpha}    
\bibliography{completepatches}{}  
\end{document}